# Comfort-as-a-Service: Designing a User-Oriented Thermal Comfort Artifact for Office Buildings

*Completed Research Paper*

**Svenja Laing**
IBM Deutschland GmbH
IBM-Allee 1, 71139 Ehningen
Germany
svenja.laing@ibm.com

**Niklas Kühl**
Karlsruhe Institute of Technology
Kaiserstr 89, 76133 Karlsruhe
Germany
kuehl@kit.edu

## Abstract

*Most people spend up to 90 % of their time indoors. However, literature in the field of facility management and related disciplines mostly focus on energy and cost saving aspects of buildings. Especially in the area of commercial buildings, only few articles take a user-centric perspective and none of them considers the subjectivity of thermal comfort. This work addresses this research gap and aims to optimize individual environmental comfort in open office environments, taking advantage of changes in modern office infrastructure and considering actual user feedback without interfering with existing systems. Based on a Design Science Research approach, we first perform a user experience testing in an exemplary corporate office building. Furthermore, we build a mechanism to gather user feedback on environmental comfort. Based on this, we build a machine learning model including different IoT data sources (e.g. building data and weather data) with an average coefficient of determination of 41.5%. Using these insights, we are able to suggest current individual comfort zones within the building and help employees to make better informed decisions on where to sit or what to wear, to feel comfortable and work productively. Therefore, we contribute to the body of knowledge by proposing a user-centric design within a cross-disciplinary context on the basis of analytical processes.*

## Introduction

In many industries, we can observe a shift towards service orientation ("servitization"). The resulting services can be manifold, e.g. from car or bike sharing services, which offer mobility as a service, over the IT industry, which is shifting from licensed software products to software as a service, up to predictive maintenance services for connected washing machines. Another area in empiric services literature are buildings and their shift from being a financial asset towards being a service provider (Pasini et al. 2016). This is one reason why smarter indoor thermal comfort for office buildings becomes more important and might become a competitive advantage. This is mainly supported by four reasons. First, most people spend up to 90 % of their time indoors (Shaikh et al. 2014). Second, the new "generation Y" of employees with a different mind-set regarding work attitude has higher comfort expectations (Parment 2009). Third, with an increased awareness of Internet of Things (IoT) technology for private homes, the overall expectations in automation and personalization increase (Breivold and Sandstroem 2015). And, fourth, the changes in office infrastructure to open office spaces with shared desk policies open up new possibilities for targeting thermal comfort challenges—which we address in this article. Knowing that buildings cause up to 40% of the worldwide energy consumption (Shaikh et al. 2014), it is not surprising that most of the related work focuses on energy and cost savings (Shaikh et al. 2014). However, only few articles regarding commercial buildings choose a user-centric perspective (Alamin et al. 2017; Freire et al. 2008; Sturzenegger et al. 2016). Even less of the articles take actual user feedback into account (Chen et al. 2015; Gupta et al. 2014; Zhao et al. 2016). None focuses on improving everyone's individual thermal comfort, but rather focus on optimizing the overall thermal comfort throughout the group. This might be because of the former office infrastructure where it was hard to account for the subjective nature of thermal comfort.

Given the current changes in office environments towards open office spaces with shared desk policy we address this research gap of user-centric, service-oriented building services for a specific use case with this article. Based on a Design Science Research (DSR) approach, we suggest, develop, implement and evaluate a thermal comfort assistant with a user-centric approach. Using this assistant, we aim for providing individualized user recommendations about the workspace the users should choose within open office spaces to optimize their thermal comfort level, thus, increasing their productivity and well-being. In comparison to related literature, we work without assumptions about thermal comfort level, e.g., norms, but instead with real, user-individual feedback. And, while the related work focuses on optimizing the building control system to satisfy most of the people, we aim for improving everyone's thermal comfort without interfering with existing systems by taking advantage of the changes in modern office environments. Therefore, we address the following research question: How can we improve every employee's individual thermal comfort in shared desk open office environments while targeting the subjective nature of thermal comfort without interfering with the building management system?

According to Hevner and Chatterjee (2010), a DSR project should cover at least three cycles of investigation, a rigor cycle (focuses on the research contribution to the knowledge base), a relevance cycle (targeting the application domain) and one or multiple design cycles (building and evaluating the research artifact). To answer the mentioned research question, we address three sub questions, one in each cycle. We investigate the knowledge base by conducting a thorough literature review based on Webster and Watson (2016) in the rigor cycle. We study the application domain in the relevance cycle doing user experience testing in an exemplary open office. Finally, we perform three design cycles, in which we build our artifact to gather user feedback, predict individual thermal comfort in certain zones and provide recommendations. In the upcoming sections, we start by describing our research design in more detail (section two). Next, we provide an overview of the related work from the rigor cycle (section three). Section four focuses on the user experience testing, the study design and related sample groups as well as a discussion of the results. Our three design cycles for gathering user feedback, data exploration and model building as well as for model deployment are described in section five. We summarize the paper in section six, regard limitations and show an outlook.

## Research Design

As an overall research design, we choose DSR, as it allows to consider the practical tasks necessary when designing IT artifacts (March and Smith 1995) and has proven to be an important and legitimate paradigm in information systems (IS) research (Gregor and Hevner 2013). Hevner and Chatterjee (2010) mention three inherent research cycles as a key part of any DSR project. A rigor, a relevance and a design cycle. The rigor cycle is one of the main differentiators between professional design and DSR. It focuses on the knowledge contribution and innovation by taking past knowledge/ research into account. The rigor cycle highlights the research contribution by combining creativity with the given knowledge base. The relevance cycle investigates the application domain of the to-be-designed artifact. As a result, the relevance cycle not only provides the requirements for the research activities, but also the acceptance criteria. The design cycle relies on the insights gained in the two previous cycles. It iterates much faster through the process of design, construction, evaluation and feedback while designing the research artifact. The gained knowledge in each of the design cycles is then lead back to the rigor and the relevance cycle.

An overview of the activities of this article is depicted in Figure 1. We start by looking into the existing knowledge base by doing a literature review based on Webster and Watson (2016). We look for thermal comfort optimization, monitoring and prediction for commercial buildings in the fields of facility management, computer science and systems engineering and control. To investigate the target domain in the relevance cycle, we conduct user experience testing based on Lewis and Rieman (1993) in an exemplary open office space. In the process of building the artifact we perform three design cycles. Thereby we follow the methodology proposed by Vaishnavi and Kuechler (2004) and the enhancements of Kuechler and Vaishnavi (2008). Since our approach is implementation heavy and we favor a clear differentiation between an abstract "suggestion" and a concrete, more programming-specific "development" this approach is well-suited for the task at hand. Vaishnavi and Kuechler generally differentiate into five process phases of a design cycle: Awareness of problem, suggestion, development, evaluation and conclusion. Each of the cycles is described in more detail in the following sections.

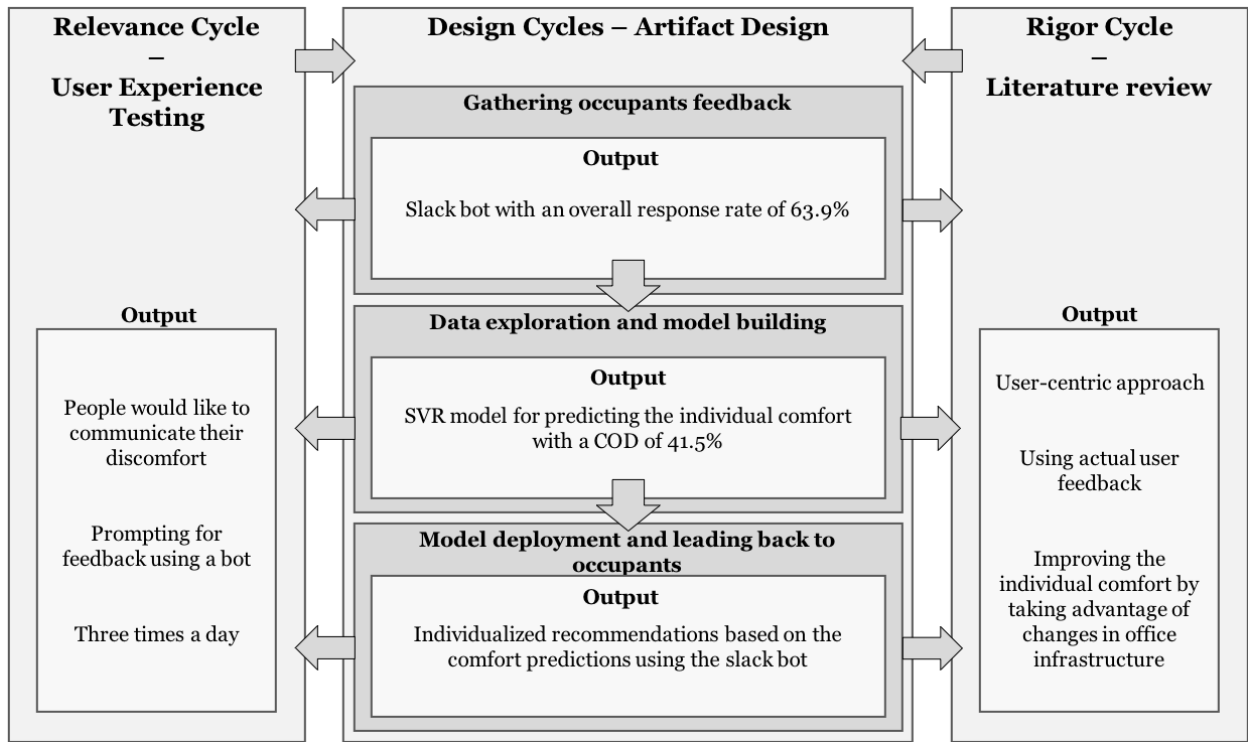

**Figure 1: Overview of the research cycles and their main outcomes**

# Related Work (Rigor Cycle)

Our goal is to optimize people's individual thermal comfort in commercial buildings. Especially in office environments we face three main challenges. First, what is the central indicator that we are trying to optimize and, second, how do we measure it? In any case, thermal comfort seems to be very subjective and individual. Which brings us to the third remaining challenge, how do we account for that subjectivity when there are several people in a certain space? We approach these questions by conducting a literature review[1] based on Webster and Watson (2016).

If we consider the field of systems engineering and control and the articles focusing on model predictive control (MPC) for buildings, the developed control strategies are mainly focused on optimized energy usage without violating certain thresholds for the occupant's thermal comfort. Those articles have different approaches on how to define thermal comfort thresholds. Some are based on certain temperature ranges (Sturzenegger et al. 2016), some take Fanger's predicted mean vote (PMV) (Fanger 1973) as measure for thermal comfort (Alamin et al. 2017; Freire et al. 2008). Others look at the actual mean vote instead of the PMV (Chen et al. 2015; Gupta et al. 2014; Zhao et al. 2016). Looking only at temperature ranges is a very easy, but maybe a too basic approach, knowing that the thermal comfort of people is dependent on more than temperature (Fanger 1973). Fanger's PMV is a seven-point scale which indicates the average thermal comfort based on air temperature, mean radiant temperature, relative air velocity, vapor pressure in ambient air, activity level and thermal resistance of clothing. It is part of the ISO 7730 standard (ISO 7730 2005) and a commonly used thermal comfort model. Nevertheless, in practice it might be challenging to get good estimates for the PMV as mean radiant temperature and relative air velocity are hard to measure and might differ between different areas in the same space (Auffenberg et al. 2015). Also, the clothing level usually gets estimated using tables provided by the ASHRAE (2009). Therefore, some of the articles take the actual mean vote of the occupants instead. Gupta et al. (2014) investigate the impact of actual user feedback with simulated user feedback. Chen et al. (2015) conduct an experiment where occupants are exposed to different environmental conditions in an experimental set up over a period of a few hours. During that time the participants are asked to report their thermal sensation (Chen et al. 2015). A longer empirical study is conducted by Zhao et al. (2016). They provided a web interface where people could give feedback about their thermal preferences over a period of three months. Their results show that people have very different perceptions of thermal sensation. This is one of the reasons why it is important to take the actual user feedback into account. On the other hand, the results of Zhao et al. (2016) show the second aspect of thermal comfort which needs to be considered; its subjective nature. The effort of most of the mentioned articles above is in optimizing the overall thermal comfort, including Zhao et al. (2016).

---

[1] Databases: IEEE, ScienceDirect, MDPI – Keywords: Smarter building, smarter building automation, commercial buildings, office buildings, indoor thermal comfort, PMV, Individual/personal thermal comfort, thermal comfort model, thermal comfort control, temperature control, smarter comfort, comfort optimization and monitoring, indoor climate control, model predictive control, data-driven comfort prediction, occupants comfort feedback, actual mean vote, actual comfort feedback

While these sources focus on energy optimization and overall thermal comfort optimization for building automation, using different thermal comfort estimates, none considers optimizing individual thermal comfort and, in general, a user-centric approach. Even though we consider the user-centric approach as essential accounting for the subjectivity of thermal comfort and its impact on productivity (ASHRAE 2009). In the combined domain of building management and the method of machine learning there are some attempts to bring people to the center of the control loop. Each of them provide different ways of human-building-interaction—mostly with mobile applications (Purdon et al. 2013; Zhao et al. 2014) or web interfaces (Daum et al. 2011). Purdon et al. (2013) base their model purely on occupants' feedback data. But nevertheless, they gather individual thermal comfort preferences for optimizing the overall thermal comfort throughout the group. In their synthesis, they show two ways of taking different thermal comfort preferences into account: Either maximizing the number of people who feel comfortable without looking at the thermal comfort of the remaining ones. Or, on the other hand, an optimization of the compromise between the extremes—so no one feels extremely uncomfortable (Purdon et al. 2013). None of those two options seem to be appropriate to address the individual thermal comfort preferences for satisfying all building occupants and ensuring thermal comfort and productivity of employees in office buildings. But the high cost pressure in facility management caused, among others, a reduction of space per person (Bedford et al. 2013). Meaning the infrastructure of office environments changed from personal offices to flexible open office spaces with shared desk policy (Bedford et al. 2013). This offers new possibilities for targeting the challenge of the subjective nature of thermal comfort.

We address this research gap by suggesting a new approach for targeting the subjective nature of thermal comfort by providing individual space recommendations based on learned thermal comfort preferences. However, unlike related articles, we aim for a user-centric approach with the main goal to provide convenience and a natural fit into the employees everyday working life without having to interfere with existing building management systems. The focus is on occupants' thermal comfort instead of energy usage, everyday usage instead of experimental or laboratory settings and decision support for the workspace choice instead of optimizing temperature set points of the building control. In the following section, we conduct a user experience testing to find a way of fulfilling these goals.

## User Experience Testing (Relevance Cycle)

In this section, we describe the activities of the relevance cycle. The relevance cycle investigates the application domain and has its main contribution in providing the requirements and the acceptance criteria for the research activities. To address the relevance cycle, we gain insights about the user experience by performing interviews. Interviews are the adequate research method as we are interested in firsthand opinions and perceptions of the building occupants to understand the application domain (Martin and Hanington 2012). Aiming for a set of requirements and evaluation criteria, we answer the following question: How can we enable human-building-interaction in a convenient way, which ensures thermal comfort of the occupants without interrupting their everyday working life?

## *Study Design and Methodology*

For the interviews, we combine structured interviews with a concurrent think-aloud protocol (Bedford et al. 2013; Gregor and Hevner 2013). In the think-aloud protocol method, participants are given certain tasks which they should complete. During task completion, they are asked to articulate what they do, think and feel (Lewis and Rieman 1993; Martin and Hanington 2012). In our case, participants have the task of giving feedback about their thermal comfort. An overview of the three main tasks the participants are asked to perform can be found in Figure 2. We start with an open question about the user's preference for human-building-interaction. Then, we design two sets of suggestions to get insights about two main characteristics for providing convenience to the occupants when giving feedback about their thermal comfort. The first set is focused on the time interval, the second on the communication medium. For both sets, mock-ups for different options are provided and the participants are asked for their thoughts. The suggested time interval goes from prompting for feedback every hour to only getting feedback when the occupant actively triggers it. For the communication medium, we provide five different options: A fixed device in an easy accessible position on the floor, receiving an email, prompting a pop-up window, interacting with a bot or a mobile application. According to the Nielsen Norman Group (Nielsen 2012), five users per task are the appropriate number of participants in a user experience testing, taking the effort versus the added value of additional users into account. After approximately 15 participants, a saturation of information is reached (Nielsen 2000). As we have two sets of suggestions—which can be considered as tasks—and enough voluntary participants available, we conduct the interviews with a sample of 16 employees from the exemplary open office space. Thereby, we ensure a mixed group regarding the different job roles, ages and gender. The participants are a sample of the target users for the to-be-designed artifact. After the interviews we perform a light weighted data analysis (Moed et al. 2012) using the transcripts of the video recordings from the interviews.

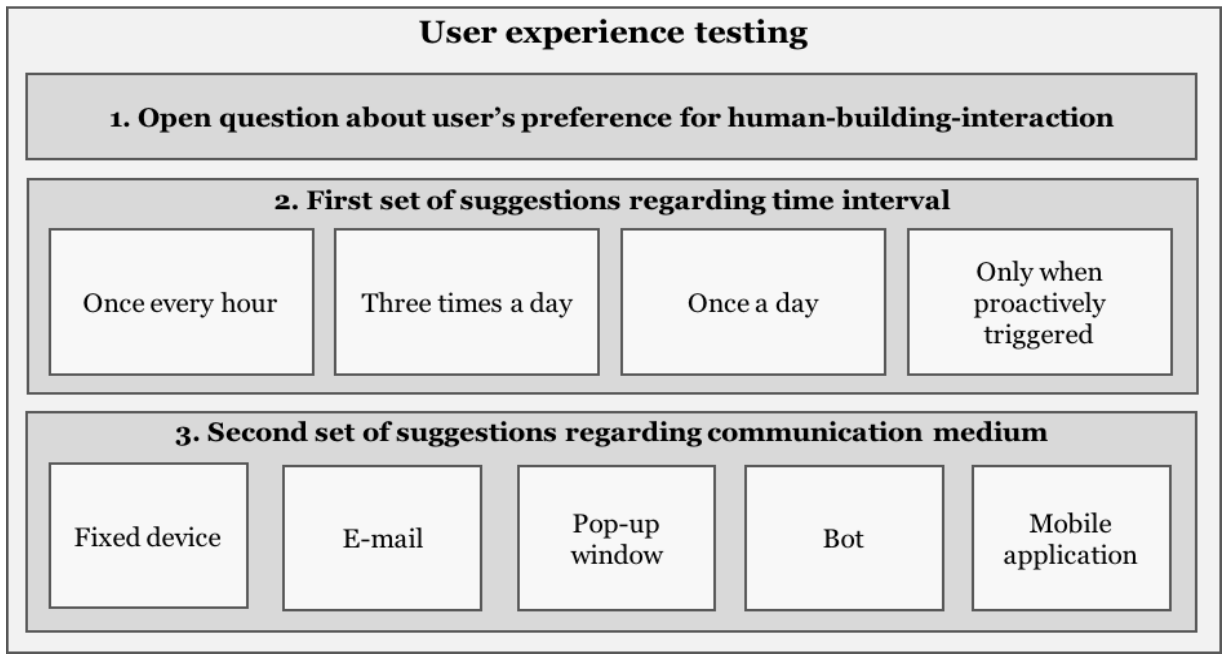

**Figure 2: The three tasks for the user experience testing, including the corresponding sets of suggestions**

## *Results*

As a result, regarding the first question about the participant's preferences for human-building-interaction, we find that most of the people would like the interaction to be via voice commands.

For the first set of suggestions regarding the time interval, nearly all the participants (94%) perceive the first option, prompting for feedback every hour, totally unacceptable and annoying. At the same time, most of the participants admit that they might not give feedback about their thermal comfort if they would have to actively trigger it unless they feel very uncomfortable. Being prompted for feedback three times a day is the most reasonable solution for the majority (94%) of the participates for a limited learning period of some weeks.

For the second set of suggestions regarding the communication medium opinions are more diverse. But most of the participants (87%) agree on the email being inappropriate. A big group of the participants (55%) like the mobile application as a long-term solution when only giving feedback if they feel uncomfortable. For giving feedback in a learning phase, when being prompt for the feedback, the majority (81%) prefers a bot.

### *Discussion*

The goal of this section is to find requirements and evaluate criteria for the research artifact—considering that the main feature of the artifact is ensuring user thermal comfort in a convenient way. The user experience testing results show that many participants like the idea of voice interaction. But thinking of the application area of an open office space, this is not feasible due to the resulting noise level. The results also highlight that asking for feedback three times a day is the right tradeoff between enough possibilities for interaction without interrupting the workflow and inconvenience the employees. In addition, the results point out two communication media that are likely to receive a high acceptance rate by the users which can be used as an evaluation criterion for convenience; a bot and a mobile application. As we aim our research asset to be adaptable in the learning phase as well as in the employees everyday working life, the bot seems to be the more appropriate solution. It can be integrated into any employee application and likely result in a solution which can be used in the learning phases as well as for the long-term human-building-interaction.

Using these insights, the actual artifact design as part of the design cycles is described in more detail in the next section.

## Artifact Design (Design Cycle)

Based on related work we studied in the rigor cycle, we found the focus points for our research to be a user-centric-approach taking advantage of the changes in office infrastructures. We aim for providing individual thermal comfort recommendations based on individually learned preferences. The results of the relevance cycle point out the design requirements and evaluation criteria. They show that a bot that prompts for feedback three times a day seems to be the most appropriate solution. Given these insights, we are targeting the following question with three constitutive design cycles: How can we build an artifact which gathers user feedback, incorporates it into a machine learning model and results in continuous, automated recommendations for occupants which increases their individual thermal comfort?

In the following, we describe each of the cycles in more detail, following the five phases of Kuechler and Vaishnavi (2008) described in the methodology section.

### *First Design Cycle – Gathering Occupants Feedback*

This first design cycle describes the process for building the part of the artifact which enables the human-building-interaction and thermal comfort reporting.

When building the artifact for gathering the feedback, we face the challenge of the feedback mechanism's need to be very simple and fast. We need to address these requirements because of our goal to provide convenience to the user, but, at the same time gather enough data for building a machine learning model in the second design cycle. Therefore, we use the response rate to evaluate whether the acceptance of the feedback mechanism is sufficient. As a baseline for the response rate we take the baseline suggested by Baruch and Holtom (2008).

**Suggestion and Development**

Considering the insights gained in the user experience testing, we suggest using a bot to gather user feedback. This bot should prompt the employees three times a day for their thermal comfort level and include location information to match with the sensor values, taken as features for the machine learning model later. Slack is a communication tool especially designed for enterprise communication (Slack 2018). It provides among many other features an open application programming interfaces (API) which allows developers to build their own Slack apps or bots (Slack 2018). Given this possibility and the fact that Slack is commonly used in the exemplary office space we are targeting, we decide to build the suggested bot as Slack bot. This Slack bot (Figure 3) prompts the users at 9:30am, 1:30pm and 4:30pm. It includes a floorplan with markers for different predefined zones. The thermal comfort level options are based on the scale of Fanger's PMV (Fanger 1973): Very cold, cold, slightly cold, good, slightly hot, hot, very hot. Users can easily respond with just two clicks, allowing a high convenience. To get an indication about the acceptance as well as the response rate, there are two additional buttons people could click on, to indicate that they are not in the testing area. One to indicate the person is just not in the area at that moment. The other to indicate, that the person is not in the office that day, so this person will not get prompted any more that particular day. In addition, the users are able to actively trigger the survey to report when they feel uncomfortable. The Slack bot got distributed to all the employees in the office space who all agreed on participating. This resulted in 36 participants over a learning period of six weeks in winter (November 28th until December 20th, 2018). Future artifacts can increase the time-span to also account for different seasons. However, the feasibility of such an artifact can also be shown with a singular season and additional models for other seasons can be built in the future.

**Figure 3: Slack bot, triggering the employees three times a day for their thermal comfort feedback**

**Evaluation**

As mentioned above, we evaluate the feedback mechanism using the response rate. We utilize the baseline suggested by Baruch and Holtom. They looked at 490 studies and found the average response rate for surveys targeting individuals is 52.7 % with a standard deviation (STD) of 20.4 (Baruch and Holtom 2008).

In our learning period of six weeks, we achieve an overall response rate of 63.9 % with a STD of 13.3 for our Slack bot, which is higher than the baseline suggested by Baruch and Holtom. Baruch and Holtom also highlight the importance of a non-response analysis, when studying response rates. If we consider the individual response rates per person, we find that we do not have any non-responses. The lowest response rate we get from one of the participants is 9.7 % and the second lowest is already 28.0 %. Seven out of the 36 participants even have a response rate higher than 100 % which indicates that they used the possibility of actively reporting their discomfort. Table 1 provides a summary of the response rate distribution overall and per person.

|  | **Mean** | **STD** | **Min** | **25%** | **50%** | **75%** | **Max** |
|---|---|---|---|---|---|---|---|
| Overall | 0.639 | 0.133 | 0.386 | 0.561 | 0.630 | 0.717 | 0.980 |
| Per person | 0.702 | 0.282 | 0.097 | 0.504 | 0.703 | 0.960 | 1.153 |

**Table 1: Slack bot response rate analysis, overall response rate and response rate per person**

Given the results of the Slack bot's response rate being clearly above the baseline, we consider the suggested solution feasible. We do not see the need for an additional iteration of this design cycle at the given stage of the research. The fact that about 20 % of the people even used the possibility for actively reporting their discomfort, highlights the acceptance and the amount of discomfort still existing in this exemplary modern office building, resulting in room for improvement.

## *Second Design Cycle – Data Exploration and Model Building*

The second design cycle is about the data exploration and model building to gather meaningful insights of the available data and allow future predictions. To start the model building, we are required to make two fundamental decisions: Frist, whether build models for every individual user or whether we build one for every zone within the space. Second, it is of importance how we define the different zones within the space, which might be used as a basis for the model and in any case as recommendation to the users in the end. The advantage of building a model for every individual user lies in the fact that we do not have to retrain all models when new users start using the artifact. However, when building a model for every individual, we need a large quantity of feedback data from every user in comparison to building one model per zone. Gathering enough individual feedback data is very time-consuming as we attach great importance to provide additional comfort to the user with as little interruptions in their everyday working life as possible. Therefore, we decide to build one model per zone. The exemplary open office space is already divided in four different zones/workspaces (Figure 4), which are the smallest units, that can be controlled individually, based on the control infrastructure of the building. For deciding on how to separate the exemplary space for the model building we consider different options, like aligning with the predefined zones or using unsupervised machine learning methods and construct the space topology based on the given data (Ahmada and Dey 2007; Baiab et al. 2011). As for further work it might be interesting to look into the interaction between the building control and the design artifact of this article, we decide to stay aligned with the predefined zones. As a result, we build four different models one for each of the predefined zones. For each model, two different machine learning algorithms are compared after tuning their parameters using a nested cross validation.

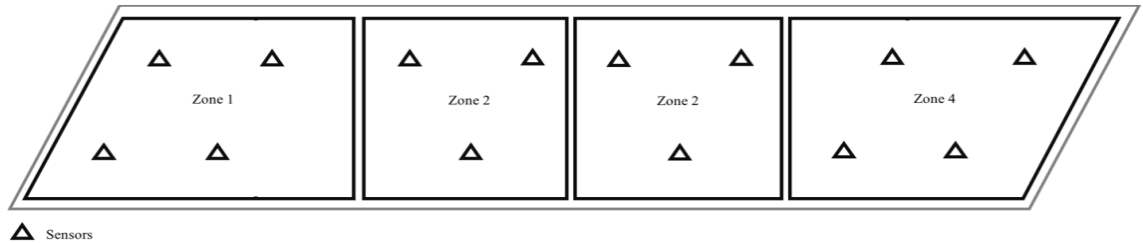

**Figure 4: Exemplary open office space, divided in four zones/ workspaces, including the sensor layout**

As most of the machine learning endeavors, this cycle is highly dependent on the quality and the amount of the available data. Due to the user-centric approach, there is a high uncertainty whether the amount and quality of the user's actual thermal comfort votes are sufficient to generate models which provide meaningful insights. To measure whether the models are meaningful or not, i.e. whether they provide added value, we take the mean model as baseline. The mean model is a prediction model, which always predicts the mean of the target variable in the considered training set (James et al. 2013). The mean model prediction results are a benchmark for the prediction performance of other regression model. Therefore, we evaluate our model in comparison to the mean model.

**Suggestion and Development**

For comparability of the received user feedback and the PVM, we base our interpretation of the occupants' thermal comfort votes on the scale of Fanger's PMV. This results in a transformation of the categorical votes (very cold, cold, slightly cold, good, slightly hot, hot, very hot) into a continuous numerical variable, between minus three and three. As the thermal comfort vote is interpreted as a continuous variable after the transformation, we suggest building a regression model for each of the predefined zones, to predict the actual individual thermal comfort levels in that zone. Using this prediction, we loop back the information to the building occupants to provide a recommendation for their individual thermal comfort zone in the building. Due to the relatively small amount of data and eventual interest in the main influencing features of the individual thermal comfort, we suggest classical machine learning algorithms like support vector regression (SVR) and random forest regression (RF) instead of deep learning algorithms.

We start building the model by collecting the data from all the different data sources for the exemplary open office space we are working with. As mentioned, this open office space is divided in four different zones/workspaces. Each of those workspaces is equipped with three to four sensors, which provide five different data types: Illuminance level, sound pressure, motion, temperature and relative humidity. For each of the sensors we calculate an approximation of the PMV and the predicted percentage of dissatisfied (PPD). In addition, we add weather data: Outside temperature, outside relative humidity and the outside UV-index. This data is merged by the nearest timestamp with the user feedback gathered from the Slack bot. As a result, we get four data sets without missing values, one for each workspace. The user feedback, so the actual thermal comfort level is the target variable, a user id, the sensor data, the calculated PMV and PPD and the weather data are features. Figure 5 shows the data structure in more detail. Based on the feedback response rate and the distribution of the participants on the floor we get different sample sizes for the workspaces (Table 2). Due to the small sample size of workspace four we focus the further work only on the workspaces one to three.

| **Workspace** | **1** | **2** | **3** | **4** |
|---|---|---|---|---|
| **Sample size** | 328 | 614 | 212 | 4 |

**Table 2: Sample sizes for the different workspaces, based on the amount of feedback responses for the workspace**

Within the data preprocessing, we use a one-hot-encoding for the user ids. As a result, every user represents a new binary feature. For all other numerical features, we perform a min-max-scaling, which transforms the feature range between zero and one (Runkler 2016). After some data exploration and visualization, we start performing a nested cross validation with a 10-fold inner cross validation for parameter tuning and model selection.

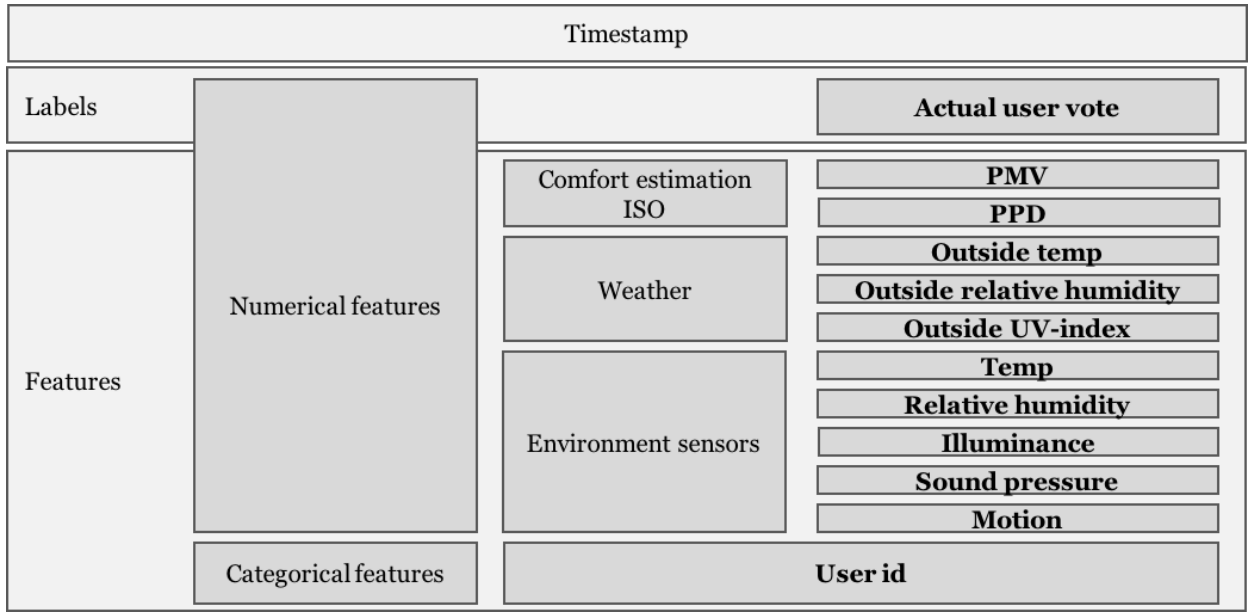

**Figure 5: Data structure of the data sets for each of the workspaces (bold: Actual data points)**

During the k-fold cross validation the data gets split into k parts (folds) of the same size. Then k iterations take place where the model is trained on k-1 folds and tested on the remaining fold (Han et al. 2011). For reasons of comparability and reproducibility, we fix and save the inner and outer fold data set for each of the five performed iterations. By performing several iterations, we get insights about robustness. The different parameters we choose during parameter tuning for the SVR and the RF can be found in Table 3.

**SVR**

| Kernel | Gamma | Penalty term (C) | Degree |
|---|---|---|---|
| RBF | [1e-3, 1e-4, 1e-5] | [1, 10, 100, 1000, 1500, 2000, 10000] | - |
| Linear | - | [1, 10, 100, 1000] | - |
| Poly | - | [1, 10, 100, 1000] | [2,3,4,5,10] |

**RF**

| Number of estimators | Maximal depth |
|---|---|
| [5, 10, 15, 20, 500, 1000] | [2, 5, 7, 9] |

**Table 3: Tuning parameters used during the nested cross validation for the SVR and the RF**

In parallel, we consider different feature selection methods like recursive feature elimination and univariate feature selection. In a final step, we perform a manual feature selection based on our domain knowledge and the results of the feature selection process of Farhan et al. (2015). Using these subsets, we perform the same nested cross validation as with the whole data set. In addition, we conduct a feature importance analysis.

**Evaluation**

We base our evaluation process on the findings of Spuler et al. (2015) regarding performance metrics for regression models. As they suggest, we use the coefficient of determination (COD) (R squared) for the parameter tuning, as we are restricted to use only one metric in this process. The COD measures the goodness of the fit and can be interpreted as the percentage of variation of the target variable explained by the prediction (Shevlyakov and Oja 2016). For the model selection and final evaluation, we use a combination of metrics. Mainly, we consider the normalized root mean squared error (NRMSE), the Pearson's correlation coefficient (CC) and the global deviation (GD) (Spuler et al. 2015) of the actual thermal comfort and the predicted thermal comfort of the user. We are aiming to minimize the NRMSE as it is based on the deviation of the prediction and the actual value. By analyzing the normalized version, we receive a better comparability between the different datasets (Spuler et al. 2015). As mentioned above, we use the mean model as benchmark for evaluating the performance of our prediction results. The mean model (James et al. 2013) is a regression model that always predicts the mean thermal comfort level of our training data. We will exemplarily discuss the results of workspace one. A more detailed summary of these can be found in Table 4 and in the plot (Figure 6).

Tuning results for the mean model

|  | COD | NRMSE | CC | GD |
|---|---|---|---|---|
| **Average** | -0.0190 | 0.2502 | 0.0 | 0.0148 |
| **Variance** | 0.0004 | 0.0004 | 0.0 | 0.0004 |

Tuning results for SVR

|  | COD | NRMSE | CC | GD |
|---|---|---|---|---|
| **Average** | 0.4146 | 0.1850 | 0.6625 | 0.0208 |
| **Variance** | 0.0005 | 0.0005 | 0.0002 | 0.0007 |

Tuning results for RF

|  | COD | NRMSE | CC | GD |
|---|---|---|---|---|
| **Average** | 0.3958 | 0.1976 | 0.6421 | 0.0089 |
| **Variance** | 0.0004 | 0.0001 | 0.0006 | 0.0001 |

**Table 4: Overview of the prediction results of the five nested cross validation iterations for workspace one, average performance metric and their variance**

As we regard the results of the nested cross validation and parameter tuning, we find the models to be robust as we get a low average variance of 0.0005 (SVR) and 0.0004 (RF) for the COD. If we compare the average results of the two models, we see that the SVR outperforms the RF for all three workspaces, even though the best parameters for each of the three workspaces are different (Table 5). For workspace one, we find the average NRMSE of the mean model to be 0.2502, while the best preforming iteration gets to an NRMSE of 0.2019. In comparison to that, the SVR gets an average NRMSE of 0.1850 and the best performing iteration has a NRMSE of 0.1503. The average NRMSE for the RF is 0.1976 while the best performing iteration is 0.1911. As both models get lower NRMSE as the mean model, independently of the iteration, this shows, that both models in any case outperform the mean model, which is what we are aiming for. On average, we achieve 24% lower NRMSE than the mean model which is a significant improvement.

However, if we look at the upper part of Figure 6 were we can see the actual user's thermal comfort vote (red) and our prediction results of the SVR (green) we notice that the model still struggles with very high (2) and very low (-3) thermal comfort votes. For that parts of the plot, we can clearly see the deviation between the red and the green line.

|  | **Best parameters** | **COD** | **NRMSE** | **CC** | **GD** |
|---|---|---|---|---|---|
| Workspace 1 | {'kernel': 'linear', 'C': 1} | 0.4412 | 0.1503 | 0.6718 | 0.0071 |
| Workspace 2 | {'kernel': 'poly', 'C': 100. 'degree': 2} | 0.5025 | 0.1197 | 0.7104 | 0.0012 |
| Workspace 3 | {'kernel': 'poly', 'C': 1000. 'degree': 2} | 0.5543 | 0.1854 | 0.7551 | 0.0137 |

**Table 5: Best performing parameter set for each of the workspaces, SVR outperforms RF in all three workspaces**

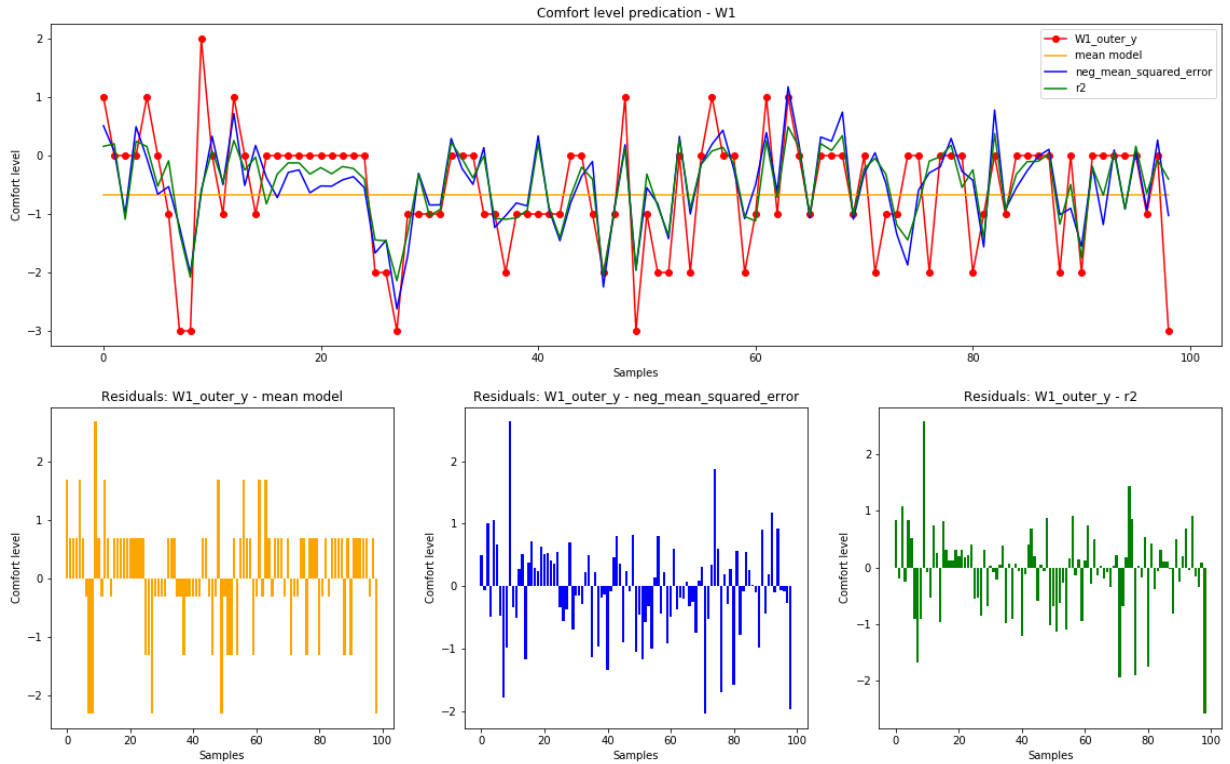

**Figure 6: Prediction results of the best performing iteration for workspace one, actual user votes (red), mean model baseline (yellow), SVR results when scoring for COD (green), when scoring for MSE (blue), below residual plots with the same color coding**

After the feature selection, the variance as well as the different performance parameters stay the same or even increase. This behavior gets confirmed by the feature importance analysis. While investigating the feature importance, we find the variance of the feature importance over all features to be very low with 0.0007. Also feature importance differs for the different workspaces and the different iterations of the nested cross validation. The only regularity we find, is that the UV-index for all workspaces and all iterations always is very low and the most important feature mostly is one of the one-hot-encoded user ids, their importance gets up to 18% at the maximum. Given these results, we decide to not look further into feature selection but stay with the original feature set.

During this design cycle, we could show the feasibility of predicting the thermal comfort level based on the actual user feedback. On average, we are about 24% better than just assuming the average. Using these insights, we are able to give recommendations for users about the workspace they should choose to optimize their thermal comfort level, which is what we are going to look into in the next design cycle.

### *Third Design Cycle – Model Deployment and Leading Back to Occupants*

The third and last design cycle is focused the model deployment and closing the loop back to our application domain and target user, the building occupant.

As we follow a user-centric approach and aim our research asset to be adaptable and valuable in the learning phase as well as in the employees everyday working life, the goal of the model deployment is to provide added value for the building occupants and, through this, increase thermal comfort and productivity of employees. To do so, the thermal comfort predictions need to be easily accessible and understandable, so that it feels like a natural fit into the user's everyday working life. The long-term acceptance rate of the design artefact by the employees will be considered to finally evaluate the third design cycle. This requires additional user research after a certain time of usage, which is not in the scope of this article and will be considered in future research.

**Suggestion and Development**

As a solution for looping back the insights into the application domain and to the target user, we suggest adding two additional informative features. One to gain insights about the average thermal comfort level of each zone in the building and the second, more valuable one, to provide individualized recommendations about thermal comfort zones, to support the desk finding process in shared desk office environments. For communication, we suggest using the same medium which was used for gathering the feedback as people already have the connection between this medium and their thermal comfort.

During the development step of this design cycle, we deploy our machine learning models from the second design cycle. To do so, we train one SVR model for each zone on the whole data set with the best parameters found in the nested cross validation (Table 5) and make it available for scoring. In addition, we add an additional slash commands to our Slack bot. This enable the employees to get individualized recommendation about their current thermal comfort zone. This trigger for recommendations takes the user's id, collects the last sensor readings, the corresponding PMV and PPD estimates and the weather forecast for each zone, then it scores each of the deployed models for the corresponding zone. According to the prediction result, the Slack bot recommends the most comfortable zone individually to the user. An exemplary recommendation is shown in Figure 7. For a better usability for the employee the prediction results are transformed back to the categorical values of the Slack bot: Comfortable (-0.5 to 0.5) slightly hot (0.5 to 1.5) or slightly cold (-1.5 to -0.5), hot (1.5 to 2.5) or cold (-2.5 to -1.5) and very hot (higher than 2.5) or very cold (lower than -2.5). The Slack bot response consists of a written recommendation including the thermal comfort class, the workspaces name and the explanation, that no better option was available when the thermal comfort class differs from comfortable. In addition, a floor plan with the highlighted thermal comfort zone is provided to help the user in quickly finding the place.

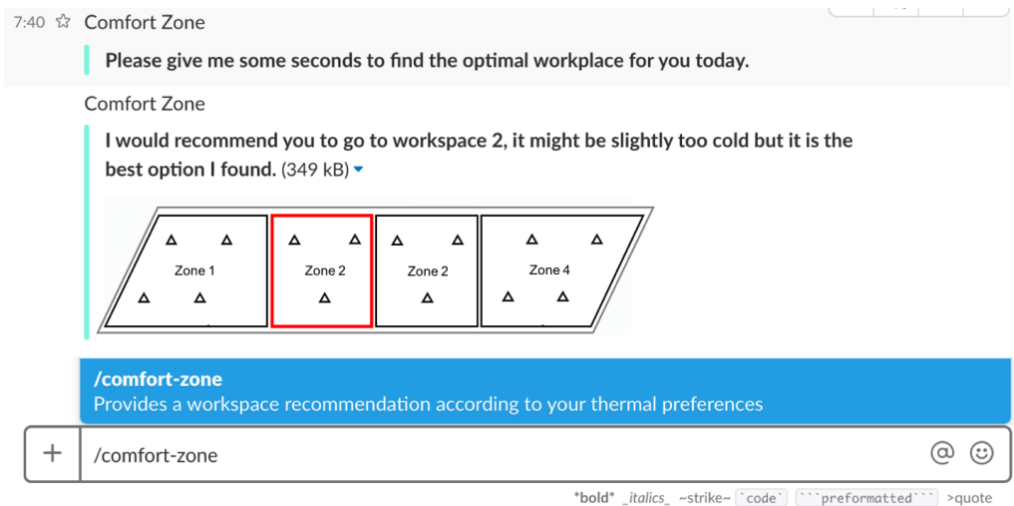

**Figure 7: Exemplary thermal comfort recommendation, slash-command triggers event, user is asked to wait for a few seconds, then the actual recommendation is displayed**

**Evaluation**

Like the focus of the user experience testing in the relevance cycle, this design cycle focuses on providing convenience to the users and improving their thermal comfort. As we have seen from the good response rates in the first design cycle, the acceptance rate is high for the Slack bot and the users are willing to use it even if there is no immediate added value during the data collection process. Therefore, we assume an even higher acceptance when users experience a directly added value—like, in this case, an immediate improvement in their thermal comfort. It is easier for people to make the mental connection between giving their feedback, getting insights and feeling the added value, when it is combined in one application. Therefore, we integrate the feedback, the models and the recommendations into one commonly used medium within the everyday work life of the employees; Slack.

First user feedback interviews reveal that users appreciate the possibility to automatically receive recommendations for choosing their workspace. However, more quantitative evaluations, including the user acceptance and their satisfaction are needed in the future. Therefore, detailed user acceptance testing based on the technology acceptance model (TAM) (Davis 1985) and its extensions (Zarmpou et al. 2012) is recommended. The adoption of the artifact should be assessed by metrics like the adoption rate or the time-to-first-action. In conjunction with the acceptance and adoption of the users, their satisfaction with the artifact is of importance and needs to be evaluated. Different possibilities on how to assess this satisfaction arise. Typical methods to achieve meaningful results are, e.g., SERVQUAL (Parasuraman et al. 1988), expectancy disconfirmation (Oliver 2014) or questionnaires (Mittal and Kamakura 2001). Future research needs to rigorously apply these methods to assess the user satisfaction of the artifact.

We show the feasibility of the direct feedback loop of the gained insights during the machine learning part to the building occupants and their immediate added value.

# Conclusion and Outlook

With this article, we aim for a user-centric approach for improving everyone's individual thermal comfort in office buildings without interfering with existing systems. Due to an increased service orientation, digitization and skills shortage, we think thermal comfort in offices is an important aspect to leverage the workspace as competitive advantage. We take advantage of the recent changes in modern office environments towards open office spaces with shared desk policy. The goal is to enable human-building-interaction for providing feedback about the individual thermal sensation in a convenient way—and use this feedback to provide individualized recommendations for thermal comfort zones within the building. In doing so, we aim to target the subjective nature of thermal comfort, improve everyone's individual thermal comfort and enable employees to be more productive. This is a novel approach compared to related work (Chen et al. 2015; Gupta et al. 2014), which are more focused on optimizing the building control and improving overall thermal comfort.

To achieve our goals, we start out with a user experience testing to find the most suitable solution for the human-building-interaction. We find this solution to be a Slack bot which is prompting for feedback. Using sensor data, weather data, the predicted mean vote (PMV) and predicted percentage of dissatisfied (PPD) and the feedback data, gathered over a six-weeks learning period, we build a regression model for each of the different workspaces/ zones in the exemplary open office space we are working with. We show the feasibility of getting meaningful insights out of a six-weeks learning period, as our models outperform the mean model. We aim for a user-centric approach. Therefore, the final step is leading back the gained insights from the prediction models to the user, by providing individualized recommendations for the most suitable thermal comfort zone. To do so, we enhance the same Slack bot further. Like this, one social information system can be used in a learning phase for gathering feedback but also afterwards for reporting discomfort and getting individualized recommendations for the current individual thermal comfort zone.

Even though we evaluate our results in an empirical setting with the exemplary open office space, the amount of people and zones is limited and might need additional research to confirm the results with additional users and zones. Also, we only consider the data gathered in winter. To further increase the artifact's usefulness and its generalizability, it would be required to add additional learning periods for different seasons, especially summer time. It would be necessary to discuss options on how different seasons can be handled within the artifact, e.g., with different designated machine learning models or with one universal model. Furthermore, the final evaluation of the third design cycle is still open. As well as the evaluation of the fulfillment of the requirement for the design artefact to be adaptable in the employees everyday working life and not only in the learning phase. In the future, it would be interesting to look into the acceptance of the application after a longer period of usage and investigate whether the employees' comfort improved. Also, an additional experimentation with integration into the building management system would be interesting. For example, it could provide explicitly different temperate comfort zones within the building according to the learned preferences of the employees. In addition, the reported discomfort could be used as alerts for facility manager. Also, an adaptation to other environmental components like noise level could be interesting to consider.

With this article, we develop an automated thermal comfort assistant which is applicable in the learning phase as well as for long term usage and addresses the subjective nature of thermal comfort. In contribution to the application domain (relevance cycle), we enable the means to improve employees' thermal comfort and satisfaction while learning about the building occupant's individual preferences. In contribution to the knowledge base (rigor cycle), we show how smart services can be utilized to improve employee's individual thermal comfort without additional expenses or difficult integrations into existing systems.

## Acknowledgements

The authors would like to thank Joern Ploennigs for his support and input given his background and broad expertise in the topic.